# Fluidic Endogenous Magnetism and Magnetic Monopole Clues from Liquid Metal Droplet Machine

Ying-Xin Zhou [1, 2], Jia-Sheng Zu [1, 2], and Jing Liu [1, 2, 3*]

1. Technical Institute of Physics and Chemistry, Chinese Academy of Sciences, Beijing 100190, China

2. School of Future Technology, University of Chinese Academy of Sciences, Beijing 100049, China

3. Department of Biomedical Engineering, School of Medicine, Tsinghua University, Beijing 100084, China

*E-mail: jliu@mail.ipc.ac.cn

**Abstract**

Magnetism and magnetic monopole are classical issues in basic physics. Conventional magnets are generally composed of rigid materials with shapes and structures unchangeable which may face challenges sometimes to answer the above questions. Here, from an alternative other than rigid magnet, we disclosed an unconventional way to generate endogenous magnetism and then construct magnetic monopole through tuning liquid metal machine. Through theoretical interpretation and conceptual experiments, we illustrated that when gallium base liquid metal in solution rotates under actuation of an external electric field, it forms an endogenous magnetic field inside which well explains the phenomenon that two such discrete metal droplets could easily fuse together, indicating their reciprocal attraction via N and S poles. Further, we conceived that the self-driving liquid metal motor was also an endogenous magnet owning the electromagnetic homology. When liquid metal in solution swallowed aluminum inside, it formed a spin motor and dynamically variable charge distribution which produced an endogenous magnetic field. This finding explains the phenomena that there often happened reflection collision and attraction fusion between running liquid metal motors which were just caused by the dynamic adjustment of their N and S polarities. Finally, we conceived that such endogenous magnet could lead to magnetic monopole and four technical routes were suggested as: 1. Matching the interior flow field of liquid metal machines; 2. Superposition between external electric effect and magnetic field; 3. Composite construction between magnetic particles and liquid metal motor; 4. Chemical ways such as via galvanic cell reaction. Overall, the fluidic endogenous magnet and the promising magnetic monopole it enabled may lead to unconventional magnetoelectric devices and applications in the near future.

# 1 Introduction

Although the magnetic field is invisible and intangible, it is a special substance that exists objectively and can be found from cells to the periphery of stars. So far, researchers have conducted in-depth studies on various existing known magnetic fields[1, 2], such as coils, earth, galaxies, etc., as shown in **Figure 1a-c**. The magnets exert forces and moments on each other through the respective magnetic fields. With the aid of the magnetic field as a medium, the object can transmit the magnetic force without contact. The study of magnetic fields is of great significance to scientific and technological progress, such as electronic cooling[3], microfluidic chip[4], graphene orientation controlling[5], medical imaging and tumor treatment[6-8], visual probe of amino acid[9], life science magnetic protein[10] and magnetotactic bacteria[11, 12] and so on.

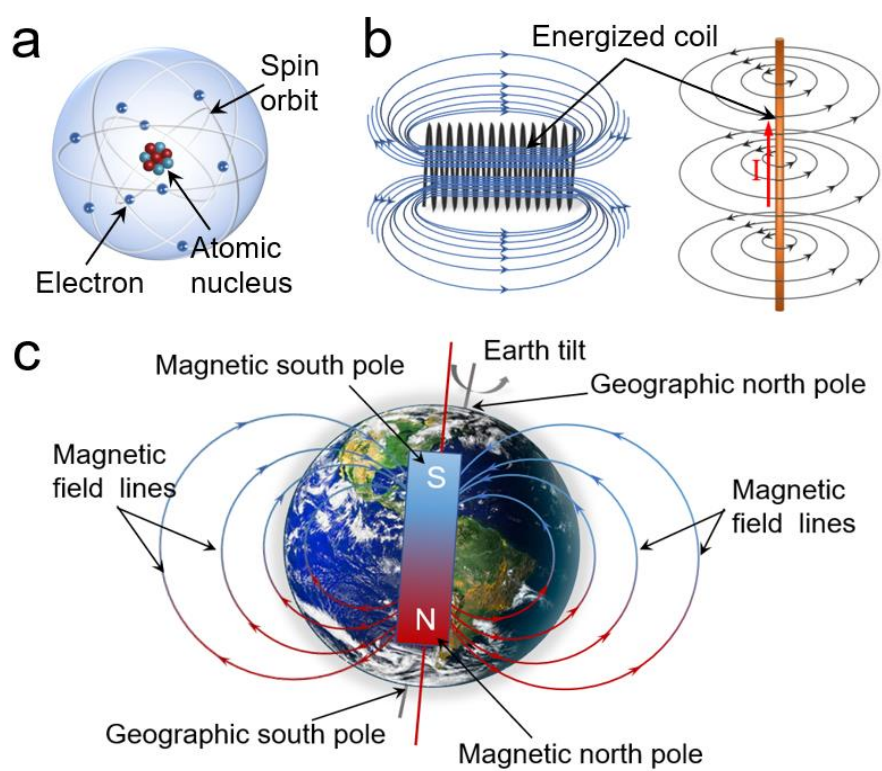

**Figure 1.** Several typical types of magnetic fields: (a) Inside the atom; (b) The energized coil; (c) The earth.

According to modern physics, the magnetic field produced by moving electric charges (**Figure 1a**) is fundamentally determined by moving electrons or moving protons. Atom is the smallest basic unit that makes up a substance, and the magnetism of a substance is the collection of the magnetism of all the atoms it contains. Inside the atom, the nucleus has spin motion, and the electrons outside the nucleus are rotating while moving around the nucleus. The above motion will generate tiny circular currents. According to Maxwell's electromagnetic theory, electricity and magnetism are induced by each other. These tiny circular currents will induce a corresponding magnetic field. Therefore, the magnetic field generated by current or point charge is the macroscopic manifestation of the magnetic field generated by a large number of moving electrons or protons.

Traditionally, from microscopic magnetic nanoparticles to macroscopic natural magnets, as well as artificially manufactured excitation coils (**Figure 1b**), permanent magnets, etc., are composed of rigid materials, and their N and S poles are located at the fixed ends of the magnet. Generally, their shape and structure cannot be changed on the eigen-scale, which makes the applications somewhat limited. In order to improve the adaptability of the magnet to different situations, people have invented the magnetic fluid functional material, which is a colloidal solution formed by surface active agent-encapsulated nanomagnetic particles dispersed in the base fluid[13], and can exhibit special magnetic and optical properties. Although the fluidity has been enhanced, the part that exhibits magnetism has not deviated from the essence of the rigid magnet. Recently, as a base fluid material for loading magnetic nanoparticles, liquid metal has been gradually developed as multi-functional materials due to its outstanding electrical and thermal conductivity and ductility. Its excellent deformability allows droplets to adapt to paths of different sizes. In particular, the chemical properties of room temperature liquid eutectic alloys such as GaIn and GaInSn are relatively stable under normal conditions. Their low toxicity and good operability allow researchers to adjust the alloy ratio to achieve various melting points and characteristics, so as to be competent for tasks that rigid materials cannot do.

Previous studies on the magnetic properties of liquid metal have mostly involved adding iron particles[14], or chemically coating a layer of nickel on the surface of the droplet[15], and then using magnets to manipulate the composite material. The orientation of the magnet will cause a change in the motion state of the droplet. Even when the position of the droplet is unknown, the magnet can attract the liquid metal to achieve rapid control. However, the story will not just end there. Here, we conceived that on a microscopic level, the spherical liquid metal droplet in fact formed a weak endogenous magnetic field when excited by its own circular current or an external current during the rotation process. This rapidly rotating liquid metal droplet not only owns the magnetic properties of rigid materials, but also flows like water. It belongs to a new kind of magnetic matter, or termed as fluidic magnet.

Further, we conceived that the discloser of the fluidic magnet still offers another important clue to answer the classical issue of magnetic monopole, one of the most intriguing scientific mysteries in nature. When people try to tackle the fundamental problem of whether the magnetic monopole exists or not, they tend to explore clues in the rigid materials or magnets which however may not always be rational in reality. The magnetic poles of such type of material are usually in a fixed position. The magnetic field lines are emitted from the N pole outside the magnet and then return to the S pole, and inside the magnet, the S pole points to the N pole, forming countless closed circuits. This fixed pattern of magnetic field distribution limits the routes to verify the actual existence of magnetic monopoles. So far, the magnetic monopoles only appear as quasi-particles in condensed matter, such as the flipping excitation of spin ice[16-18], and the similar structure produced by the vortex of super-cold rubidium atoms in Bose-Einstein condensate (BEC)[19]. However, we realized that as a variable fluidic conductor, the liquid metal machine generates a constantly changing endogenous magnetic field in the random spin motion. When an external electric or magnetic field is superimposed, or compounding with magnetic particles, or just artificially modifying the magnetic field through internal chemical reactions, even much diverse magnetic field configurations or behaviors will be generated. This is fundamentally different from the conventional rigid magnetic material with fixed-poles. This article is dedicated to present a new conceptual fluidic magnet and provide several possible technical routes towards finding the magnetic monopoles.

## 2 Experimental Evidences of Fluidic Endogenous Magnetism from Liquid Metal Machine
### 2.1 Basic Properties of Conductive Fluidic Metal

At the macro level, the flexibility or rigidity of material has an important impact on its application. In recent years, with the development of material science, various unique effects of soft matter have been gradually discovered and utilized. Particularly, room temperature liquid metal owns many favorable properties, including large surface tension, ideal flexibility, high conductivity, low toxicity, etc. From Table 1[20], it can be seen that the typical liquid metals generally own very similar fluidity with water while their thermal conductivities are dozens of times higher than the later. Especially, the excellent electric conductivity of such matters molds them into intrinsically conductive fluid. Basically, electrically conductive solution such as aqueous NaOH, NaCl, etc. may also display similar behavior with liquid metal. But considering the huge conductivity of liquid metal, such fluid will particularly play dominate role in developing liquid magnet and is therefore the core of current analysis.

**Table 1** Typical physical properties of Ga-based liquid metal with those of other liquids[20]

| Composition | Ga | $Ga_{75.5}In_{24.5}$ | $Ga_{67}In_{20.5}Sn_{12.5}$ | $Ga_{61}In_{25}Sn_{13}Zn_1$ | Hg | Water |
|---|---|---|---|---|---|---|
| Melting Point (°C) | 29.8 | 15.5 | 10.5 | 7.6 | −38.8 | 0 |
| Boiling Point (°C) | 2204 | 2000 | >1300 | >900 | 883 | 100 |
| Density (kg/m$^3$) | 6080 | 6280 | 6360 | 6500 | 13530 | 1000 |
| Electrical conductivity (/Ω/m) | $3.7 \times 10^6$ | $3.4 \times 10^6$ | $3.1 \times 10^6$ | $2.8 \times 10^6$ | $1.0 \times 10^6$ | - |
| Thermal conductivity (W/m/K) | 29.4 | 26 | 16.5 | - | 8.34 | 0.6 |
| Viscosity (m$^2$/s) | $3.24 \times 10^{-7}$ | $2.7 \times 10^{-7}$ | $2.98 \times 10^{-7}$ | $7.11 \times 10^{-8}$ | $13.5 \times 10^{-7}$ | $1.0 \times 10^{-6}$ |
| Surface tension (N/m) | 0.7 | 0.624 | 0.533 | 0.5 | 0.5 | 0.072 |
| Sound speed (m/s) | 2860 | 2740 | 2730 | 2700 | 1450 | 1497 |
| Water compatibility | Insoluble | Insoluble | Insoluble | Insoluble | Insoluble | - |

As gradually realized by recent studies, the excellent characteristics of liquid metal make it available in the fields of fluidics (**Figure 2a**), printed circuits (**Figure 2b**), flexible sensors, transformable machine (**Figure 2c**), self-driving motors (**Figure 2d**), and the base carrier fluid for magnetic particles (**Figure 2e**), etc. The present research focuses on the electromagnetic effects of such fluidic conductive material. Particularly, the shape and motion of liquid metal have great variability and controllability, the magnetic field generated by it will have rather rich possibilities, which appears more competent in some occasions. In this sense, we can come up to a generalized matter state, which can be termed as electromagnefluid, indicating the material that simultaneously consisted of electronics, magnet and fluid inside together.

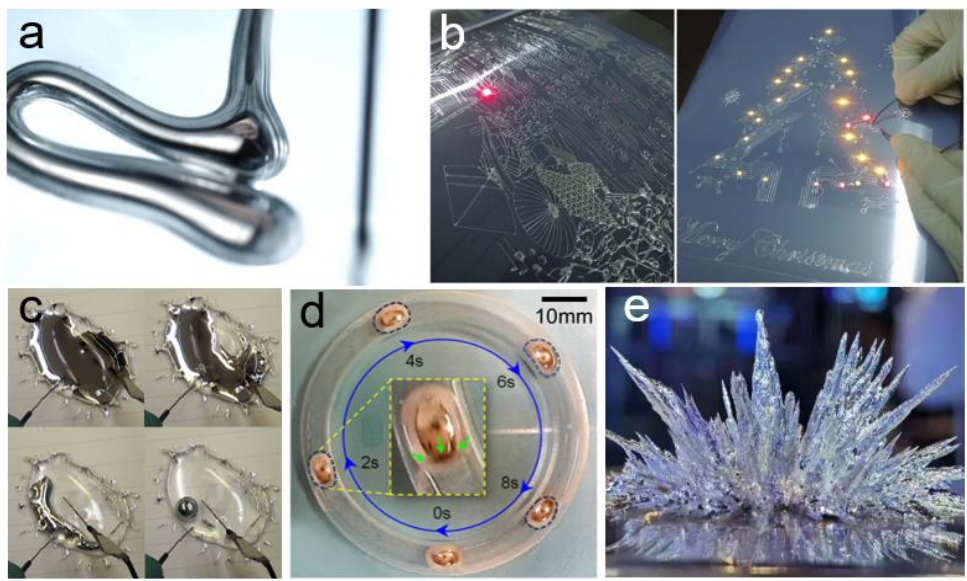

**Figure 2.** Typical material features of liquid metal which may serve to make endogenous magnetic fields: (a) Fluidic behavior; (b) Conductive electronic ink[21]; (c) Morphologically transformable machine[22]; (d) Self-fueled motor[23]; (e) Loadable by magnetic particles[24].

**2.2 Magnet Induced from Liquid Metal Energized Coil**

Liquid metal has good flexibility and conductivity, and can replace traditional rigid coils, making the moving parts of the electromagnetic actuator more flexible, thereby improving the ability to deal with complex situations. At present, there are two main methods to make liquid metal electromagnetic coils. One is to use optical mask technology to lithography microchannels on the PDMS substrate, and then inject liquid metal into it[25]. In this way, the liquid metal is still in fluidic form (**Figure 3a**).

Another method is to use a mask with a specific shape to cover the PDMS substrate, and uniformly print the liquid metal on the PDMS film through a liquid metal spray gun (**Figure 3b**)[26]. Compared to the first method, this direct printing method of liquid metal has a shorter production cycle and simpler operation. In this case, the liquid metal exists on the PDMS in a semi-solid form, that is, the part in contact with air is oxidized to solid, but the liquid metal inside still has the possibility of flowing behaviors.

In the magnetic driving device, the magnet was placed above or below the liquid metal electromagnetic coil (**Figure 3c**), the magnetic field lines passed through the electromagnetic coil. When an alternating current was applied, the Lorentz force $F$ was generated in the coil:

$$\mathbf{F} = \int I d\mathbf{l} \times \mathbf{B} \qquad (1)$$

where, $dU$ as indicated in **Figure 3d** is the potential difference between the two ends of the micro segment $dl$. Macroscopically, the Lorentz force generated by the magnetic field component parallel to the coil was embodied as driving the coil to move closer or away from the magnet.

For the first injection method, since the flow channel is in a sealed state and the operating temperature is higher than the freezing point of the liquid metal, the alloy in the flow channel is always liquid. Under the action of an external electric field, in addition to inducing a magnetic field, the micro segment would also generate an endogenous magnetic field under an alternating current. This segment had weak attractive or repulsive interaction with the placed magnet and other micro segments. However, this force was too weak, only the device action dominated by the Lorentz force of the coil could be observed.

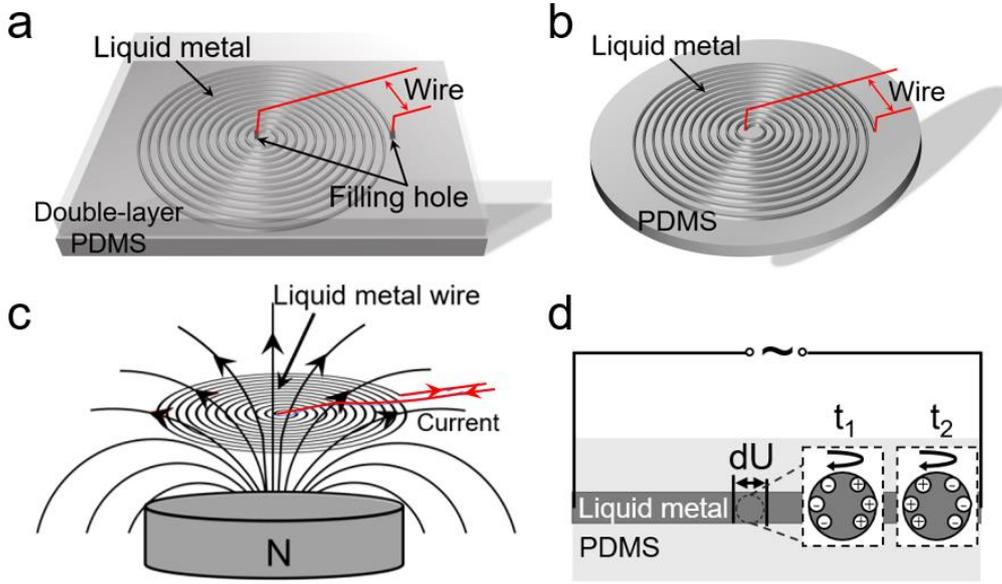

**Figure 3.** Schematic diagrams of liquid metal coil: (a) Method of injection; (b) Method of printing; (c) Magnetic field around the coil; (d) Partial enlarged view of the charge movement and rotation of the liquid metal in the coil.

**2.3 Endogenous Magnet Generated by Electrically Controlled Liquid Metal Machine**

Applying an electric field to the liquid metal can induce its transformation, movement or rotation, and its motion state depends on the direction and strength of the applied electric field, the contact with the electrode, and the surrounding solution environment, respectively[22]. To illustrate the basic principle to generate endogenous magnet by electrically controlled liquid metal machine, we designed a cylindrical channel with a smooth surface, and put a spherical liquid metal droplet in the electrolyte-filled solution (**Figure 4**). After applying an external field, the liquid metal immediately responded, rotating and moving towards the positive electrode. The original spherical droplet was stretched, taking the advancing direction of the droplet as the head, it was observed that the tail had a slight deformation. When the applied electric field was large enough, the droplets were dragged instantaneously (**Figure 4a-d**), and even separated into two small droplets, as shown in **Figure 4d**.

Due to the difference between the physical properties of the liquid metal and the electrolyte solution, the electric field would be stepped at the two-phase contact interface, thereby generating electrical stress. At the same time, GaIn liquid metal could react with NaOH solution to generate $[Ga(OH)_4]^-$, $[Ga(OH)_4]^-$ would be specifically adsorbed to the Ga surface, that is, in the Helmholtz layer, and the surface of Ga was negatively charged due to the abundance of reactive electrons, and the diffusion layer was positively charged, forming an electric double layer (EDL). According to Lippmann's equation, there was a connection between surface tension and potential difference[23]:

$$\gamma = \gamma_0 - \frac{1}{2}cV^2 \qquad (2)$$

where, $\gamma$ is the surface tension, $c$ is the capacitance per unit area of EDL, $V$ is the potential difference of EDL, and $\gamma_0$ is the maximum surface tension when $V=0$.

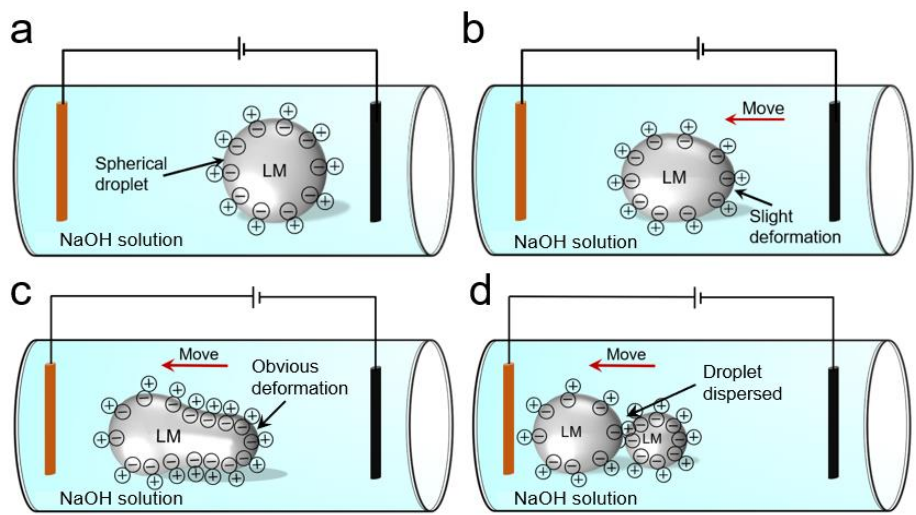

**Figure 4.** The liquid metal droplet continues to move to the electrode under the electric field: (a) The droplet was a regular spherical shape; (b) The droplet moved to the positive electrode with slight deformation; (c) The droplet had obvious head and tail; (d) Droplet dispersed.

EDL makes the charge distribution of two droplets similar. Using an electrode to drag a droplet and approach another in the solution, when the distance between the two was small enough, they would quickly merge. This internal rotation caused by an external electric field was similar to the spontaneous rotation of the droplet after swallowing aluminum, and we will explain this in detail in section 2.4.

In the system composed of liquid metal and water, it could be found that the influence of liquid metal on the surrounding water is rather evident under the action of an external electric field. The system layout is shown in **Figure 5a**, placing a spherical liquid metal droplet in water and extending a pair of electrodes into the surrounding solution. When the current was applied, the liquid metal sphere started to rotate, and two eddies appeared in the surrounding solution, and they kept spinning with the rotation of the liquid metal sphere. Therefore, the external electric field could induce the liquid metal to rotate, causing the real-time change of the electronic charge distribution on the surface of the droplet. This moving charge would cause the droplet to generate an endogenous magnetic field, as shown in **Figure 5b**.

Because the flow direction of the droplet surface was along the electric field gradient, when applying an alternating current, the direction of the electric field changed periodically, and the flow state of the liquid metal would also change accordingly, as shown in **Figure 5c**. Before and after the external electric field was applied, the electric field on the liquid metal-electrolyte interface would change (**Figure 5d**). As the alternating frequency increased, the vortex current formed inside could be stronger, and the magnetic field generated inside will be strengthened at this time. It should be noted that such endogenous magnetic field was a manifestation of electricity and magnetism on a microscopic level, and its intensity was much smaller than the magnetic field induced by the change of the external electric field.

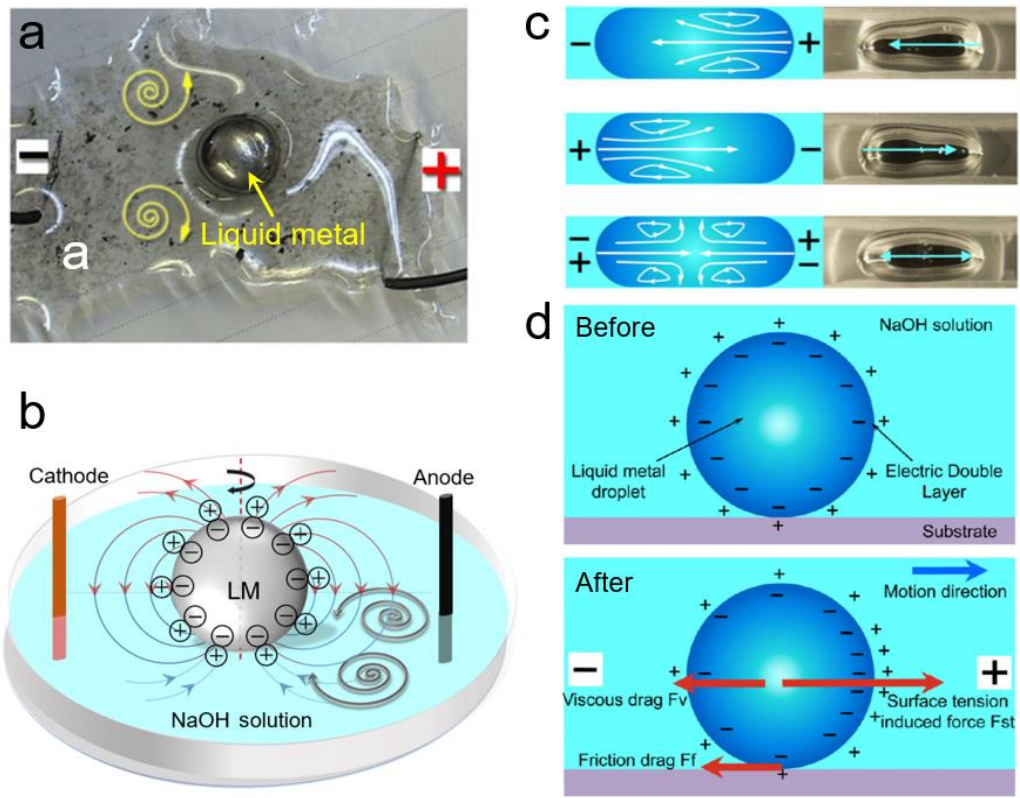

**Figure 5.** The rotation and transformation of liquid metal in the solution under an external electric field: (a) Experimental case of rotating liquid metal droplet under direct current[22]; (b) Schematic for rotating liquid metal droplet under direct current; (c) Resonant oscillation phenomenon of liquid metal droplet under matching alternating field; (d) Electric field on the liquid metal-electrolyte interface before and after applying external electric field[27].

**2.4 Endogenous Magnetic Field Generated from Self-Driving Liquid Metal Motor**

According to our former discovery, small liquid metal droplets owns intriguing self-driving capability after fueled with aluminum[28], and can automatically converge or diverge[29], which is a very unconventional feature that traditional rigid machines do not own. We placed a spherical liquid metal droplet in the NaOH solution and added some aluminum foil, as shown in **Figure 6a**. According to Rebinder's effect, after a period of time, the aluminum foil was completely infiltrated and corroded by the liquid metal, the oxide layer on the aluminum surface was destroyed, causing Al to activate, triggering a redox reaction, and the electrons from the aluminum interior preferentially deoxidated the oxidized gallium near the Al. The charge distribution of the EDL had changed, resulting in a potential gradient and asymmetric surface tension on the surface of the droplet[30]. According to the Young-Laplace equation, the pressure difference *p* between the solution and the liquid metal droplet can be expressed as:

$$p = \gamma \cdot \frac{2}{r} \qquad (3)$$

where, $1/r$ is the curvature of the droplet surface[31].

The liquid metal sphere moved randomly in the solution and was accompanied by its own rapid rotation, as shown in **Figure 6b**. And their lifetime lasted for more than 1 hour without any other external energy[23]. According to our former measurement, this self-fueled liquid metal motor generated electrical voltage and current between its surface and the surrounding electrolyte (**Figure 6c**). It is this dynamically variable electrical field leads to the generation of an endogenous magnetism inside the liquid metal motor.

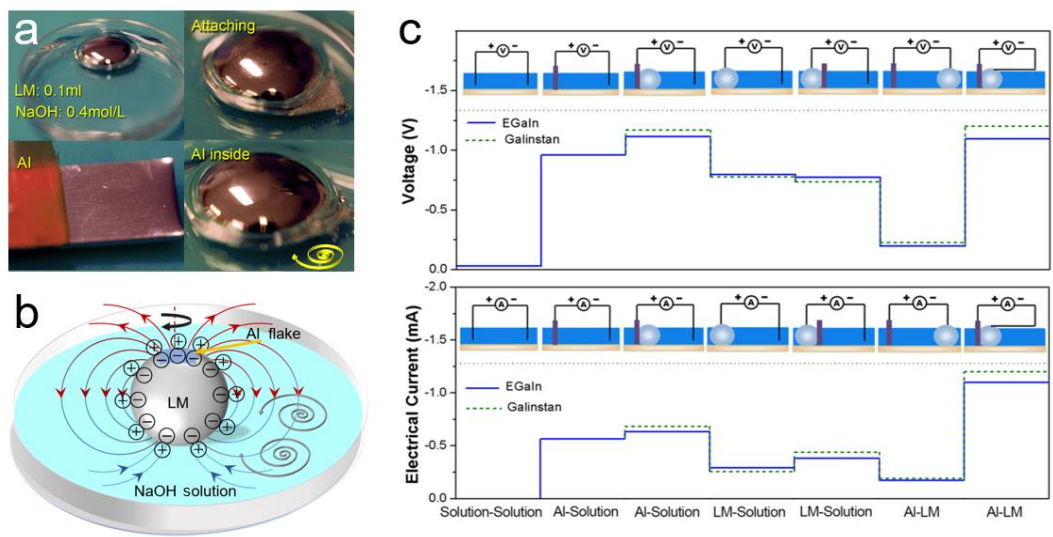

**Figure 6.** The rotation and transformation of liquid metal in the solution after swallowing aluminum: (a) Experimental case of rotating liquid metal droplet after swallowing aluminum[32]; (b) Schematic for rotating liquid metal droplet after swallowing aluminum and generated magnetic field; (c) Measured average voltage and electrical current on self-fueled liquid metal and electrolyte[23].

In addition, there was a galvanic reaction composed of liquid metal, aluminum foil and electrolyte solution, which accelerated the speed of electrochemical reaction. After being corroded by GaIn alloy, aluminum aggregated or dispersed on the surface, then it was uniformly distributed in the liquid metal in the form of small particles, and hydrogen was generated by the electrochemical reaction (Equation 4). The average voltage and current between the liquid metal and the electrolyte could be measured by Avometer[23], as shown in **Figure 6c**. The gas released from the interface also pushed the liquid metal forward, forming a self-driving motor. When the size of the droplet was larger, the specific surface area would reduce and the electrochemical reaction sites were limited, resulting in less gas generated, and the driving force of the gas does not have an obvious effect on the large droplet. Therefore, the self-driving of liquid metal in a large volume mainly depended on the tension gradient, and the effect of the latter was negligible.

$$2Al + 2NaOH + 2H_2O = 2NaAlO_2 + 3H_2 \uparrow \tag{4}$$

Next, we explore the mergence of rotating droplets (**Figure 7**). In the solution, two liquid metal droplets rotated inside after swallowing aluminum, changing the flow line of the surrounding fluid[32], and when the distance between the two is small enough, they will quickly merge into one. It could be seen from section 2.3 that the surface charge distribution of the droplets changed in the electrolyte and formed an electric double layer. This structural similarity promoted the mergence of the droplets. Here, we proposed that the rotation inside the droplet not only enabled itself and the surrounding flow field to maintain a state of motion, but also had the ability to excite an endogenous magnetic field. During the rotation process, through adaptive adjustment, it induced a magnetic field that attracted each other. The two droplets were brought closer and merged into one, as shown in

**Figure 7a-d**. Over the process, various endogenous magnetic fields were induced inside the liquid metal (**Figure 7e-h**). This mechanism allows discrete droplet machines to be quickly assembled, which has profound significance for self-assembly machine with internal trigger motion.

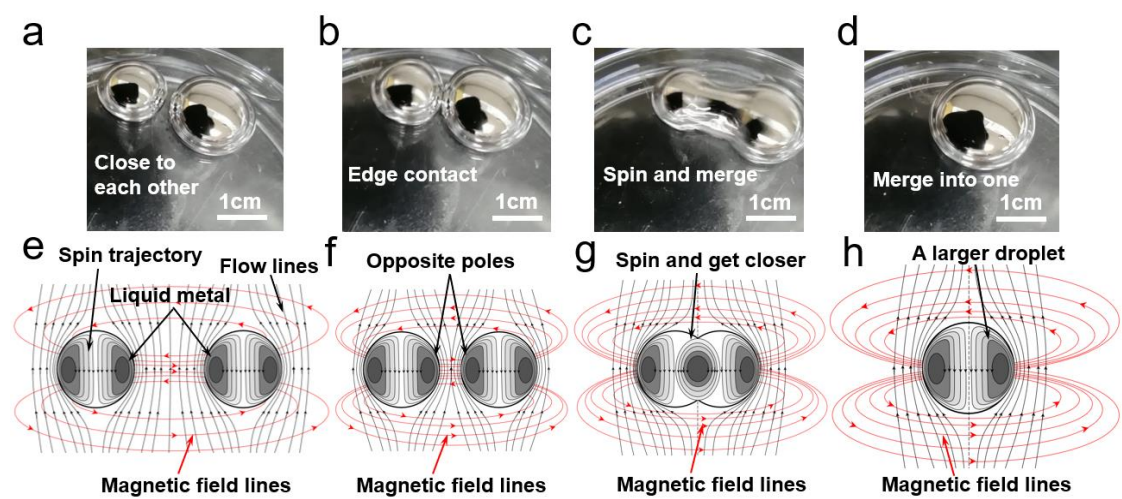

**Figure 7.** Droplets mergence and induced flow and magnetic field: (a-d) Experimental picture of the mergence process of two droplets; (e-h) Spin trajectory of the liquid metal droplets and the surrounding flow lines.

## 2.5 Liquid Metal Spin Superimposed on an External Magnetic Field

Adding aluminum in the liquid metal can form tiny motors in the NaOH solution, but the movement is random and lacks of a specific direction and speed[32]. If this kind of random movement can be controlled, the liquid metal self-driving motor can be used in more occasions, such as transformable intelligent robots, precise drug delivery, detectors/sensors, etc. We placed a permanent magnet under the petri dish to introduce a magnetic field to adjust the random movement of the motor. The $GaIn_{10}$ liquid metal motor had a significant group effect at the boundary of the permanent magnet. Over the time, the motor group gathered near the boundary of the magnet, and then rebounded after a short stay. Tan *et al.*[33] suggested that this peculiar phenomenon was related to the magnetic flux density of the bottom magnet, and the magnetic intensity was zero at the boundary of the magnet. At first, these droplet motors were attracted by the magnet, which leaded to aggregation. The higher magnetic intensity on the side of the magnet prevented the motor from passing, while the peak of the magnetic field on the side away from the magnet was lower. The droplet motor selectively tended to the low peak position, thus limiting the motor's range of motion.

Based on our former experimental video [33], we tracked the motion path of a single droplet motor, and carried out a more in-depth interpretation on the interaction mechanism between the liquid metal and the permanent magnet. Here, we proposed a conjecture that the spin of liquid metal generated a weak magnetic field on a microscopic level. Since the motor was small enough and in a solution environment, the spin drive of the motor could more clearly show the characteristics related to the magnet. The magnetic field generated by the spin is non-directional, and the droplet exhibited the same or opposite magnetic poles as the magnet at the boundary of the magnetic field, resulting in the effect of being repelled or attracted.

We have studied the effect of the external magnetic field on the moving liquid metal droplet in the above phenomenon. From the present analysis, it is known that the rotating liquid metal droplet will generate a magnetic field, and there will be negatively charged free electrons flowing through it. The magnetic field intensity of a permanent magnet around it is $H$ and the magnetic induction intensity is $B$. Then the liquid metal sphere will be subjected to the Lorentz force of permanent magnet induction intensity on the charged sphere, and the force of magnetic field intensity on the sphere's own magnetic field at the same time. Firstly, the Lorentz force on the sphere is analyzed. Assuming that the moving speed of the liquid metal sphere is $v$ and the overall charge is $Q$ at a certain time, the liquid metal sphere can be regarded as a large charged particle. Then the force $F_1$ at this time can be obtained from the Lorentz force formula in classical electromagnetics:

$$\boldsymbol{F_1} = Q\boldsymbol{v} \times \boldsymbol{B} \tag{5}$$

The stress of the liquid metal sphere in the magnetic field intensity at the same time is analyzed by using the equivalent magnetic charge theory which can be available in classical text books. Assuming that the magnetization of the liquid metal sphere is $M_p$ and the magnetization in the solution is $M_s$, then the equivalent surface magnetic charge density on the surface of the liquid metal sphere $\kappa$ is:

$$\kappa = \mu_0 (\boldsymbol{M_p} - \boldsymbol{M_s}) \cdot \boldsymbol{n} \tag{6}$$

where, $\mu_0$ is the vacuum permeability and $n$ is the normal vector of the surface of the liquid metal sphere.

By integrating the surface $S$ of the sphere, it can be obtained that the magnet product force of the sphere under the magnetic field intensity is:

$$\boldsymbol{F_2} = \oiint_S \kappa \boldsymbol{H}\, ds = \mu_0 \oiint_S (\boldsymbol{M_p} - \boldsymbol{M_s}) \cdot \boldsymbol{n}\, \boldsymbol{H}\, ds \tag{7}$$

Therefore, the total force of the applied magnetic field on the liquid metal sphere in this system is:

$$F_m = F_1 + F_2 \tag{8}$$

For a small-volume liquid metal motor, as shown in **Figure 8(a-d)**, the magnetic field strength was the smallest near the magnetic field boundary, and the magnetic force on the droplet motor was weak. Therefore, the motor oscillated in a small range nearby. The force was converted into the momentum of the drop, making the momentum continuously accumulated. When the motor moved away from the magnetic field, the magnetic pole facing the magnet was the same as the magnet, which produced a repulsive force. When it reached a certain value, it could break away from the restraint of the magnetic field boundary, and it appeared to be bounced off by the magnet on the macroscopic view. Due to the increase of the magnetic field along the diameter of the magnet, the droplet far away from the magnet would be in a larger magnetic flux density space, the magnetic force would become more obvious, and the momentum would further increase. Although the magnetic poles generated by the endogenous magnetic field might be attracted by the permanent magnet at the next moment, the attractive force could not pull back the large momentum droplet away from the magnet at this time, so the droplet eventually moved away from the magnet.

However, large-volume motors were not as sensitive to weak magnetic fields at the boundary of magnets as small-volume motors, as shown in **Figure 8(e-h)**. When the large-volume motor happened to approach the permanent magnet with the same pole, it would be hindered by the external magnetic field. In the process of approaching the center of the magnetic field, the motor's potential energy continued to accumulate and the momentum gradually decreased. Under the continuous accumulation of reverse acceleration, the liquid metal motor began to move away from the magnet (**Figure 8i**), and the accumulated potential energy turn back to its own momentum. When the speed reached a certain level, it could break away from the restraint of the magnetic field boundary.

Based on the above discussion, we suggested that when the motor had the same or opposite magnetism as the bottom permanent magnet, it was bounced or attracted (**Figure 8j**), resulting in the macroscopic effect of the motor group rotating, oscillating, gathering and bouncing off the boundary of the magnet, and this effect was affected by the volume of the motor.

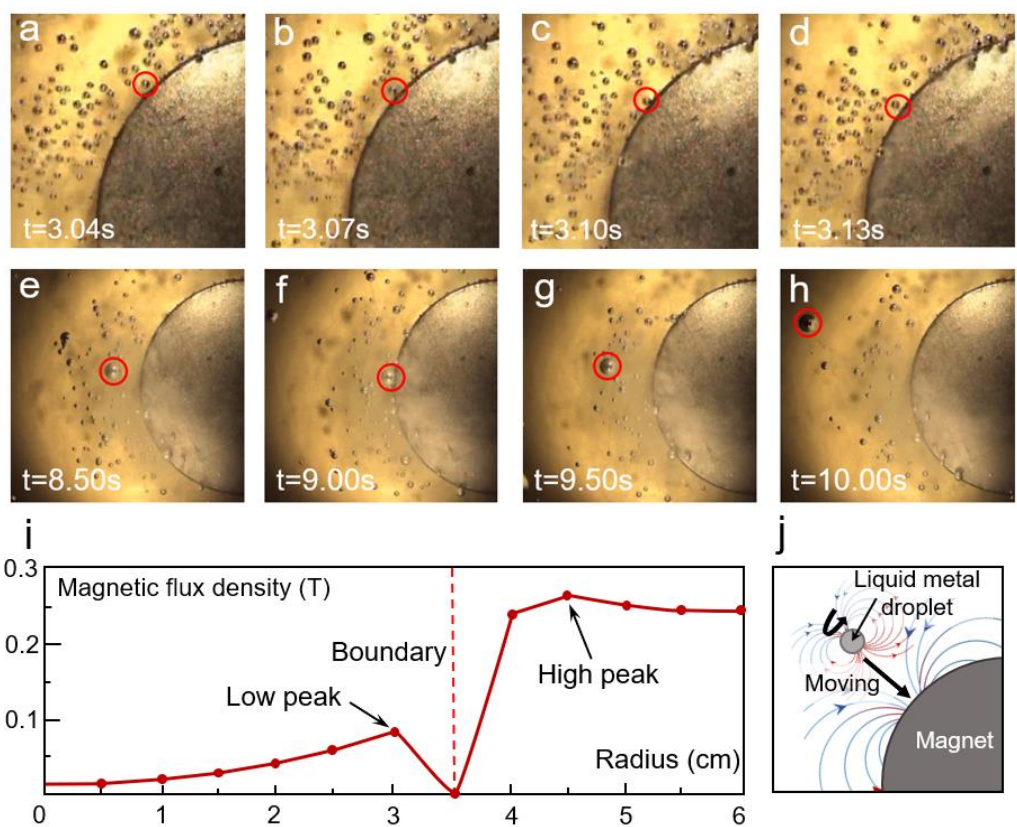

**Figure 8.** The motions of the liquid metal motor under the action of magnetism: (a-d) The position of the small liquid metal motor; (e-h) The position of the large liquid metal motor; (Data from[33]) (i) Magnetic flux density along the diameter of the magnet; (j) Schematic diagram of magnetic interaction between liquid metal motor and magnet.

## 2.6 Magnetic Field Generated from Liquid Metal under Chemical Environment

Liquid metal reacts with other metals to form a galvanic cell, which is also a form of causing self-rotation. Since liquid metal is easily oxidized in the air and forms an oxide layer on the surface, which is not conducive to the observation and analysis of the surface and internal movement. Usually, we put the liquid metal in an acid or alkaline solution for operation to remove the surface oxide film, and on this basis, build a galvanic battery with other metals.

Previous studies have revealed the Marangoni effect caused by gallium-copper galvanic corrosion couple[34], that is, the phenomenon in which the tension gradient between the liquid interface moves the mass. Here, liquid metal was used as the anode, Ga was oxidized to $Ga^{3+}$, and copper was adopted as the anode. The cathode had a corrosion potential between ($Ga/Ga^{3+}$) and ($H/H^+$), so $H^+$ was reduced to $H_2$ on the cathode (Equation 9, 10). Therefore, the essence of the liquid metal Marangoni phenomenon was the galvanic corrosion.

$$Ga \rightarrow Ga^{3+} + 3e^-  \quad (9)$$
$$H^+ + 2e^- \rightarrow H_2 \quad (10)$$

Ga reacted with HCl solution to produce gas at the liquid interface. By tracking the movement trajectory of the bubble, the flow state of the liquid metal could be obtained, as shown in **Figure 9a**, the liquid metal on the left was in contact with the copper sheet, and the flow lines of liquid gallium was marked in yellow. Here, the liquid metal participated in the reaction as a part of the galvanic battery, and the internal rotation could make the charge continue to flow through it (**Figure 9b, c**), which also had the conditions to excite the endogenous magnetic field.

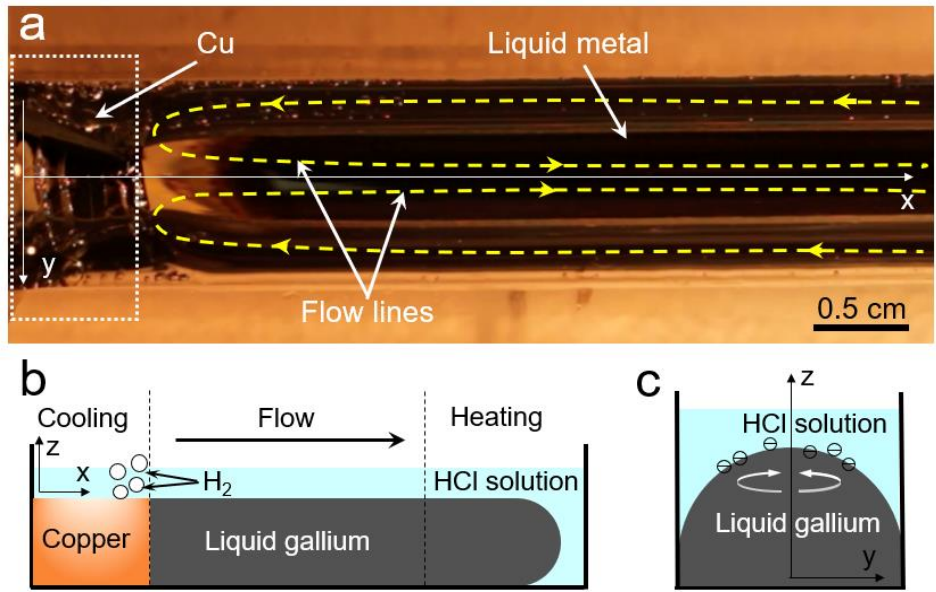

**Figure 9.** Gallium-copper galvanic corrosion: (a) Superficial solution streamline [34]; (b) The x-z plan view of galvanic corrosion system in the channel; (c) The y-z plan view of the liquid gallium and HCl solution in the channel.

In addition to the above-mentioned characteristics, liquid metal also has unique reversibility, that is, via synthetically chemical-electrical mechanism (SCHEME) to control its structure[35]. As shown in **Figure 10a**, two platinum electrodes were inserted into the liquid metal and electrolyte. According to our former research, when direct current was applied, an oxide layer was formed on the gallium surface at the anode, and the surface tension decreased. The originally spherical droplet appeared to be spread out (**Figure 10b**), and the surface area increased, even reaching five times the original size. When the applied electric field was removed, the gallium oxide on the surface chemically reacted with NaOH solution, and the surface tension of the droplet increased and returned to the spherical state, as shown in **Figure 10c, d**. This reversible SCHEME was affected by many factors, such as current intensity, electrode spacing, liquid metal volume, electrolyte concentration and so on.

In this system that combines chemical dissolution and electrochemical oxidation, liquid metal was both a conductor and a reactant. In the process of continuous expansion and contraction, the change of surface tension and the redistribution of charge could be achieved by controlling the external electric field. In the process, the internal oscillations were synergistic with the movement of electric charges, and the electric charges of this kind of motions were also related to the dynamic magnetic field. Therefore, the internal chemical mechanism of liquid metal also had the possibility of generating an endogenous magnetic field.

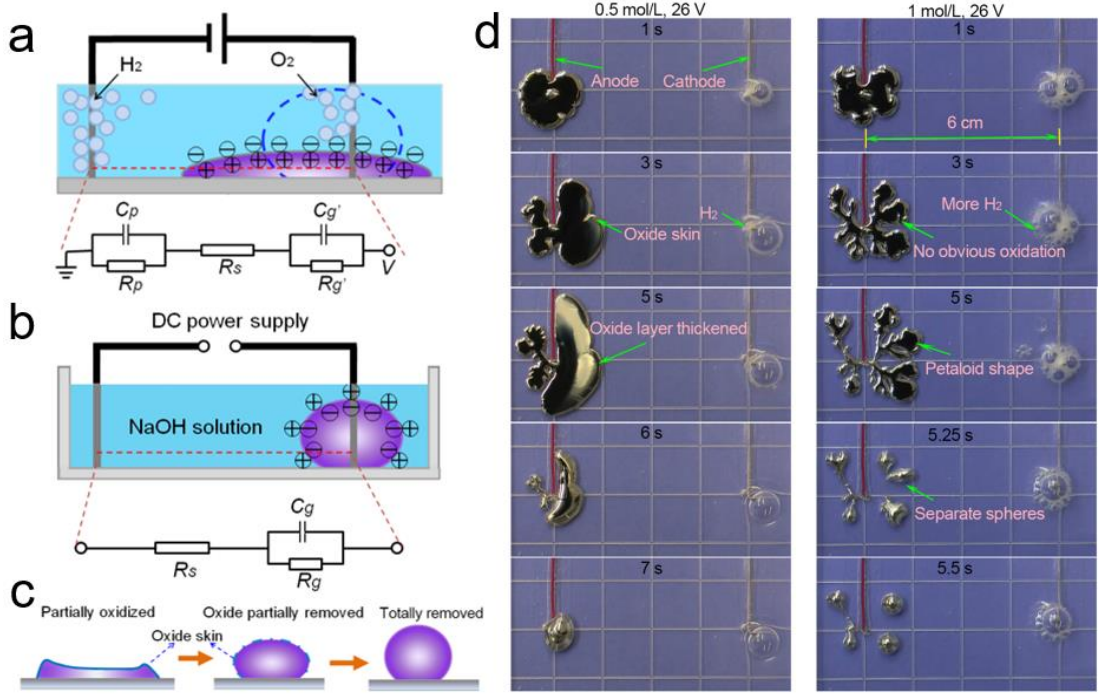

**Figure 10.** Reversible shape and charge redistribution mechanism of liquid metal [35]: (a) Experimental schematic diagram and equivalent circuit after applying DC power; (b) Experimental schematic diagram and equivalent circuit without external power supply; (c) The effect of surface oxide layer on droplet morphology; (d) The change process of the droplet shape when the external power supply was removed after 5s.

## 3. Theoretical Interpretation
### 3.1 Generation of Endogenous Magnetic Field from Liquid Metal Machines

When applied an external electric field or reacts with other metals to form a galvanic cell, the liquid metal acts as a conductor, there is current passing through its surface, the inside of the droplet rotates at the same time. From the above experiments, we have observed that liquid metal can generate endogenous magnetic fields through rotation. Next, based on some basic principles available in classical electromagnetics textbooks, we will conduct theoretical analysis and formula derivation of this peculiar phenomenon, and try to make a more in-depth explanation. For the physical properties of the liquid metal spherical droplet, its density $\rho$, electrical conductivity $\sigma$, and magnetic permeability $\mu_0$ are all constants. According to Maxwell's electromagnetic equation, the electromagnetic equation describing the space around the spherical droplet of liquid metal can be obtained as:

$$\nabla \times \boldsymbol{H} = \frac{\partial \boldsymbol{D}}{\partial t} + \boldsymbol{J}_e \tag{11}$$

$$\nabla \times \boldsymbol{E} = -\frac{\partial \boldsymbol{B}}{\partial t} \tag{12}$$

$$\nabla \cdot \boldsymbol{B} = 0 \tag{13}$$

$$\nabla \cdot \boldsymbol{D} = \rho_e \tag{14}$$

And its physical property equations are:
$$\mathbf{B} = \mu_0 \mathbf{H} \tag{15}$$
$$\mathbf{J} = \sigma \mathbf{E} \tag{16}$$
$$\mathbf{D} = \varepsilon \mathbf{E} \tag{17}$$

where, $H$ is the magnetic field strength, the unit is A·m$^{-1}$. $E$ is the electric field strength, the unit is V·m$^{-1}$. $B$ is the magnetic induction, the unit is T. $J$ is the current density, the unit is A·m$^{-2}$; $D$ is the electric displacement, the unit is C·m$^{-1}$; $\varepsilon$ is the dielectric constant.

Firstly, we analyze the liquid metal droplet in rotating state after swallowing aluminum foil, as shown in **Figure 11a, b**. The liquid metal droplet, aluminum foil and electrolyte solution together constitute a short-circuit galvanic cell system, with aluminum foil as the cathode and liquid metal as the anode. For the rotating liquid metal droplet, its internal current will be composed of two parts, the first part is the galvanic current $I_1$ produced by the electrochemical reaction of galvanic cell, and the second part is the rotating current $I_2$ produced by the charge moving in the rotating process. The current density of the liquid metal droplet is also composed of these two parts, named $J_1$ and $J_2$ respectively, and the synthetic magnetic field of liquid metal droplet is generated under these two different currents.

We define the angular velocity of the liquid metal droplet as $\omega$, and suppose that when there is current flowing inside, the direction of the current is parallel to the rotation axis of the liquid metal droplet in dynamic equilibrium. The radius of the sphere of the liquid metal droplet is 'a', the center of the sphere is the origin, and the axis of rotation is the polar axis to establish a spherical coordinate system, $r$ is the distance from any point to the center of the sphere. Assuming that $Q_1$ is the total charge passing through the maximum cross section (radius=a) of the liquid metal droplet within time $t$, the current passing through the maximum cross section can be equivalent to:

$$I_1 = \frac{Q_1}{t} \tag{18}$$

During time $t$, the current density flows on any circular cross section perpendicular to the current direction on the sphere is as follows:

$$\mathbf{J_1}(r) = \frac{I_1}{s_1} \tag{19}$$

$$s_1 = \pi(r \sin\theta)^2 \tag{20}$$

where, $s_1$ is the area of circular cross section, and there is $0 < r \leqslant a$.

For the second part, the liquid metal droplet will also produce current $I_2$ in the process of rotating. Assuming that in the process of constant rotation, the charge amount of the liquid metal droplet is $Q_2$, then the volume charge density is:

$$q = \frac{Q_2}{\frac{4}{3}\pi a^3} \tag{21}$$

The corresponding current density is:

$$\mathbf{J_2}(r) = \frac{Q_2}{\frac{4}{3}\pi R^3} \cdot \boldsymbol{\omega} \times \mathbf{r} = \frac{3Q_2}{4\pi R^3} \boldsymbol{\omega} \times \mathbf{r} \tag{22}$$

Under steady conditions, using the formula of the vector potential generated by the current in the classical theory of electromagnetics, the vector potential of any point $A$ ($r_1$, $\theta$, $\varphi$) outside the liquid metal droplet can be obtained as:

$$\mathbf{A_1}(\mathbf{r_1}) = \frac{\mu_0}{4\pi} \int_V \frac{J_1(r) + J_2(r)}{|r_1 - r|} dV \tag{23}$$

where, $|r_1-r|$ is the distance between point $r_1$ $(r_1, \theta_1, \varphi_1)$ and point $r$ $(r, \theta, \varphi)$.

According to the basic equation under steady electric fields, the galvanic current and the rotating current flowing through the liquid metal droplets can be obtained together at any point $A$ outside the body. The resulting expression of the total magnetic induction is:

$$\boldsymbol{B_1(r_1) = \nabla \times A_1(r_1)} \tag{24}$$

Next, we analyze the liquid metal sphere rotating under a constant external electric field, as shown in **Figure 11c, d**. This is similar to the previous analysis. The only difference is that the liquid metal sphere droplet does not have the current generated by the galvanic reaction, but has an external conduction current under the external electric field. When an electric field is applied, the liquid metal droplet in the electrolyte solution will not only conduct current, but also form a rotating current. In this case, it is also assumed that $I_3$ is the conduction current flowing through the largest cross-section (r=a) of the sphere, then the area current density flows on the circular cross-section perpendicular to the current direction is:

$$\boldsymbol{J_3(r) = \frac{I_3}{s_2}} \tag{25}$$

Similarly, it is assumed that liquid metal droplets will also generate a rotating current $I_4$. During the constant rotating process, it always has a charge of $Q_4$, so the charge density is:

$$q = \frac{Q_4}{\frac{4}{3}\pi a^3} \tag{26}$$

Its current density is:

$$\boldsymbol{J_4(r) = \frac{Q_4}{\frac{4}{3}\pi R^3} \cdot \omega \times r = \frac{3Q_4}{4\pi R^3} \omega \times r} \tag{27}$$

Similar to the above analysis, the vector potential of any point $B$ $(r_2, \theta_2, \varphi_2)$ outside the body of the liquid metal droplet can be obtained by using the same current to generate the vector potential formula as:

$$\boldsymbol{A_2(r_2) = \frac{\mu_0}{4\pi} \int_V \frac{J_3(r)+J_4(r)}{|r_2-r|} dV} \tag{28}$$

where, $|r_2-r|$ is the distance between point $r_2$ $(r_2, \theta_2, \varphi_2)$ and point $r$ $(r, \theta, \varphi)$.

Then the liquid metal droplet under external electric field is integrated by the conduction current and the rotating current, and the total magnetic induction intensity generated by any point $B$ outside the body is expressed as:

$$\boldsymbol{B_2(r_2) = \nabla \times A_2(r_2)} \tag{29}$$

Through the analysis and calculation of the above two motion phenomena of the liquid metal sphere droplet, it can be seen that when the liquid metal droplet is rotating in the solution and there is current passing through it, the sphere itself will also generate a rotating circular current due to the rotating motion. The superimposition of these two currents will generate a magnetic field together. At the same time, due to the instability of the rotation of the liquid metal spherical droplet on the plane, its rotational direction will suddenly change within a period of time, resulting in a change in the direction of the magnetic field.

What needs to be pointed out here is that the spherical liquid metal droplets that rotated by swallowing aluminum foil generate a specific magnetic field, which is essentially different from the magnetic fluid widely reported in the past. This kind of liquid metal is still in the form of full fluid. It can adjust the rotation speed, volume and current in the rotation process, and can also possiblly be applied to various research applications that require flexible magnetic fluid by virtue of the characteristics of full fluid.

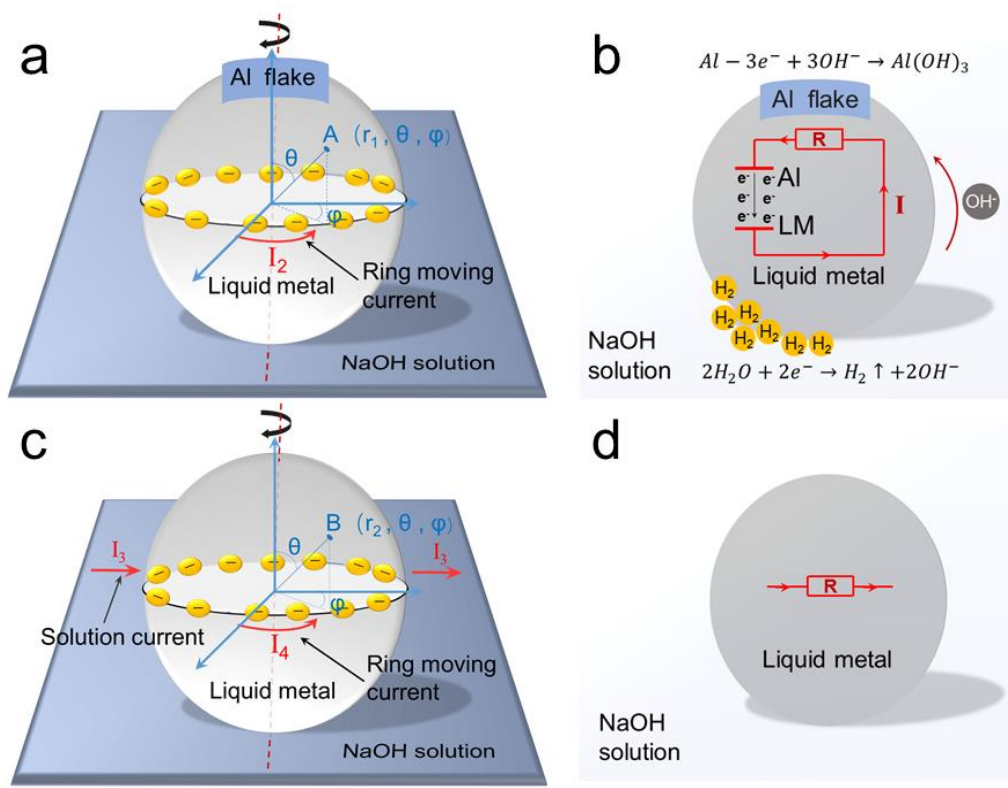

**Figure 11.** Schematic diagram of the movement mechanism and equivalent circuit of surface current in NaOH solution: (a) and (b) liquid metal sphere swallowing aluminum; (c) and (d) Under external electric field.

**3.2 Technical Routes to Realize Magnetic Monopole from Liquid Metal Machines**

Most of the liquid metal magnetoelectric devices and magnetic drive structures reported in the past used an external electric field to change the overall magnetic field of the structure, thereby causing the device to act. For example, a layer of spiral GaIn alloy was printed on PDMS to obtain a flexible electromagnetic driver[26]. When alternating current was applied to both ends of the coil, the coil would generate a magnetic field, which was attracted and repelled by the magnets on both sides according to the alternating frequency of the current, which could simulate jellyfish swimming, fish tail swinging and other actions as made in the present authors' lab. There were also studies using the electromagnetic interaction between a magnet and a Galinstan liquid metal coil to generate sound waves and made a retractable dynamic acoustic device[36].

However, the spin motion of liquid metal can excite the endogenous magnetic field in essence. Because the two magnetic poles of a traditional rigid materials are relatively fixed, it is difficult to find a matching substance when searching for magnetic monopoles, and it is even more difficult to find a case through experimental observation. Considering the aforementioned four basic ways of exciting the magnetic field with liquid metal, we proposed several possible methods of artificially synthesizing a liquid metal sphere similar to a magnetic monopolar state.

**Route 1: Matching the interior flow field of liquid metal droplet machines.** As shown in **Figure 12a**, while the liquid metal sphere rotates itself after swallowing the aluminum foil, there may be vortex-shaped small spheres inside which move opposite to the outer fluid. This unique nested structure makes it possible to artificially synthesize magnetic monopoles. When the outer sphere rotates around the axis, the inner four small spheres perform their respective circular motions in opposite directions, and then through micro-scale manipulation, such as regulating the position and speed of the rotation, the magnetic effect of the N pole or S pole generated in the outer area and the magnetic poles generated by the four small spheres in the interior are mutually offset, so that the entire liquid metal sphere exhibits only a single magnetic pole.

**Route 2: Superposition between external electric effect and magnetic field.** Secondly, an external field superposition method can be adopted. As shown in **Figure 12b**, a tiny permanent magnet is inserted into the bottom end of the liquid metal sphere. When the liquid metal sphere is spinning, the permanent magnet is fixed, so that the magnetic field of the permanent magnet and the magnetic field generated by the spin of the liquid metal sphere are superimposed. Since the permanent magnet is at one end of the liquid metal sphere, the endogenous magnetic field excited by the spin interacts with the magnetic field of the permanent magnet. When the two cancel out at a certain moment, the remaining magnetic field is located at the end far away from the permanent magnet. At this time, the liquid metal sphere will exhibit such un-cancelled magnetism.

**Route 3: Composite construction between magnetic particles and liquid metal motor.** In addition, adding appropriate magnetic nanoparticles into the liquid metal sphere is also a possible method, as shown in **Figure 12c**. Since the ferromagnetic particles are very tiny, when they are uniformly mixed with the liquid metal spheres, the spinning liquid metal spheres will also generate many tiny magnetic fields. The endogenous magnetic field can also interact with the ferromagnetic particles added. Proper control of the number, shape and position of magnetic nanoparticles can also make the synthetic magnetic field of the liquid metal sphere exhibit a certain unipolar characteristic.

**Route 4: Chemical ways such via galvanic cell reaction between liquid metal motor and substrate or foreign substances.** Finally, chemical reaction can generally be a possible option, as shown in **Figure 12d**. As mentioned above, when the copper contacts the surface of the liquid metal to form an electrode pair, a galvanic battery system that can undergo electrochemical reactions is formed in an alkaline solution. When the current flow through the liquid metal, the droplet spins under the influence of the surface tension gradient. In such an environment where two different metals are in contact with each other, it is allowed to simultaneously manipulate another metal substrate (like copper) and a spinning liquid metal sphere, so that the sphere produces a synthetic magnetic field with an apparent single pole characteristic.

Overall, in the above four different methods of constructing the single magnetic pole, the control of the spin vortex inside the liquid metal is a certain challenge due to the changeable rotation state of the droplet, and the other three methods have certain feasibility. Therefore, it is possible to synthesize a magnetic substance that acts as a source or sink of magnetic field lines by using the spin characteristics of the liquid metal sphere and through artificial control, which is expected to provide supporting evidence for the actual existence of magnetic monopoles in the coming time.

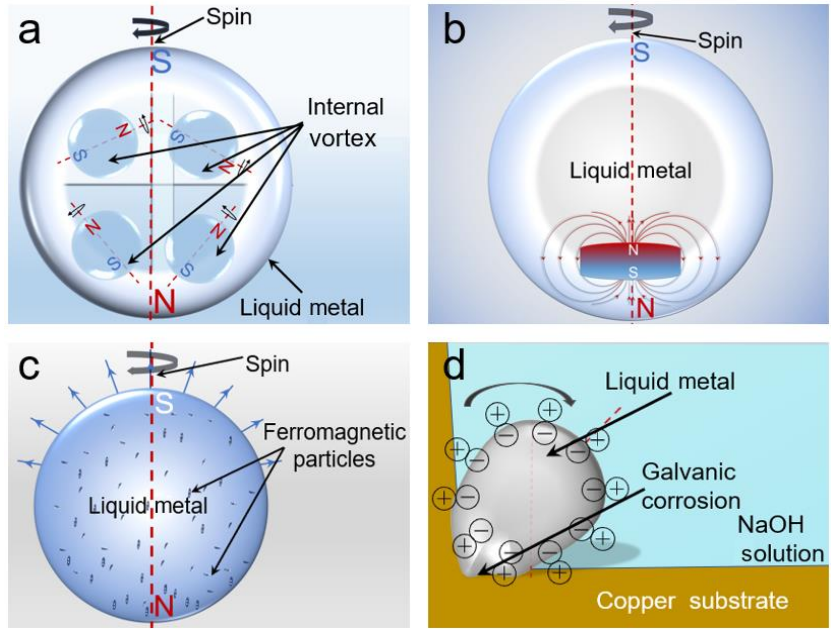

**Figure 12.** Four methods of forming a liquid metal sphere with a combined magnetic monopole field: (a) Internal vortex interaction; (b) Combined magnetic field with external magnetic field; (c) Adding magnetic particles; (d) Chemical reaction pathway.

## 4 Discussion
### 4.1 Revisit of Magnetic Monopole Theory

Under active vacuum conditions, Maxwell's equations do not satisfy the invariance of electromagnetic duality, that is, electrons are regarded as the source and sink of electric field lines, implying that particles with basic magnetic charges are related to the emission of magnetic field lines. Therefore, Dirac proposed the magnetic monopole theory in 1931 to resolve this problem[37]. A magnetic monopole is a magnetic substance with only a single magnetic pole of N or S pole in theoretical physics, and its magnetic line of induction is similar to the electric field lines of a point charge, as shown in **Figure 13a**. Dirac used a mathematical formula to guess its existence by analyzing the phase uncertainty of the wave function of a quantum system. It should be pointed out that if the magnetic monopole really exists, the electric charge and magnetic charge must be quantized in quantum mechanics. In modern physics, charge quantization has become an important theory, and this theory has also promoted the process of researchers looking for magnetic monopoles.

In order to find magnetic monopoles, researchers have conducted a lot of explorations for decades. A well-known theory related to magnetic monopoles is the Grand Unified Theory (GUTs), in which strong interaction forces, weak interaction forces and electromagnetic forces can be unified into a normative interaction. Unlike elementary particles, a magnetic monopole is a regional energy called solitary wave, the size and mass of a magnetic monopole can be estimated through unified field theory, and the quality of a magnetic monopole is approximately about $10^{16}$ GeV[38, 39]. At the same time, the internal space freedom of the gauge theory makes the magnetic monopole have a certain topological structure and provides a guarantee for stability. Although GUTs is dedicated to unifying the interaction forces between microscopic particles, it has not been finally verified, nor can it solve problems such as the "excess" of magnetic monopoles[40]. In addition, String theory, M theory, etc. also conjectured the existence of magnetic monopoles.

## 4.2 Extension of Magnetic Monopole Concepts

Except for the classical way, some physicists turned to strictly mathematical approaches to find our possible clues towards magnetic monopole. For example, Yang *et al.*[41] introduced the overlapping area of the sphere to construct a dual coordinate system to eliminate singular chords, and proved that any curl of the two vector potentials could reasonably give the magnetic field of monopole, and gave a magnetic monopole solution in the overall description of the gauge field, that is, Wu-Yang magnetic monopole. Abrikosov[42] proposed that the quantum state on the sphere of graphene was related to monopole harmonics. Under certain operators, keeping the magnetic monopole at the center of the sphere unchanged, rotating the sphere would produce a part related to spin, and this part was caused by a magnetic monopole.

On the basis of rich theories, some researchers have tried to explore the existence of magnetic monopoles through experimental means. For example, by mixing tiny magnetic spins[43], as shown in **Figure 13b**, or artificially constructing a magnetic monopole at the end of a nano-magnetic needle[44]. Fang *et al.*[45] found that there was an anomalous Hall effect in ferromagnetic crystals, and only the hypothesis of the existence of magnetic monopoles could explain this phenomenon. This result might serve as indirect evidence for the existence of magnetic monopoles in the momentum space of the crystal.

Because spin ice exhibits unique magnetic field characteristics, some studies have found analogues of magnetic monopoles in strange spin ice[16-18]. This is due to the dipolar magnetic excitation caused by s spin flipping that produces free defects distributed in the crystal lattice (**Figure 13c**), thus exhibiting the properties of a magnetic monopole[46], which has magnetic charges and electric dipoles[47], but it cannot be separated from the material[18]. Dusad *et al.*[48] used a superconducting quantum interference device (SQUID) to detect the quantization jump of the magnetic flux of the $Dy_2Ti_2O_7$ crystal, and found that the magnetic noise generated by the magnetic monopole could be heard by humans after amplification. This is the more intuitive observation of magnetic monopoles in spin ice so far.

In the field of condensed matter physics, theoretically, the spin structure of BEC could be manipulated with point-like topological defects through an external magnetic field[49]. On this basis, Ray *et al.*[50] used cold atoms to generate Raman transitions from lasers of different frequencies under a magnetic field gradient to simulate the effect of the magnetic field. Since the magnetic spin arrangement of rubidium atoms was controllable, the magnetic spins were arranged in the form of large vortices, and the middle part would produce the effect of synthesizing magnetic monopoles. At this time, the spin direction of the atom could point to a certain point in space at the same time, which also proved the basic quantum characteristics of the magnetic monopole, as shown in **Figure 13d**.

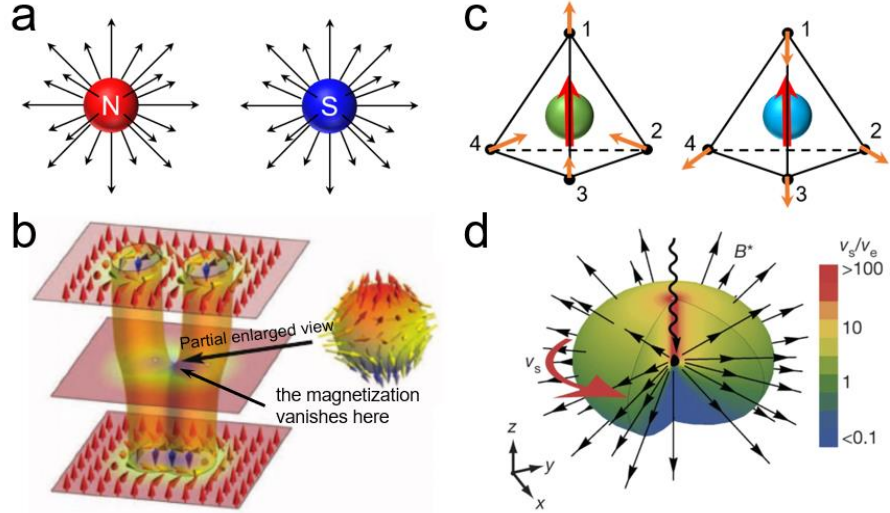

**Figure 13.** Hypothesis and simulation of magnetic monopole: (a) Conceptual diagram of magnetic monopoles; (b) Sketch of a magnetic configuration describing the merging of two skyrmions[43]; (c) Schematic diagram of spin-ice excited state; (d) Magnetic monopole model composed of cold atom condensed state[50].

**4.3 Experimental Challenges and New Fluidic Magnet Opportunity**

Although abundant evidence for the existence of analogues of magnetic monopoles has been found, the direct observation of Dirac monopoles in the medium of a quantum field has not yet achieved a breakthrough. For decades, with the support of GUTs, String theory, M-theory and other theories, researchers have even searched for relevant magnetic monopole trajectories in particle accelerators and lunar geotechnical sampling. But researchers have not yet observed its actual existence by experimental means. However, liquid metal combines the properties of electrical conductors, magnets, and fluids, which endorsed it rather profound possibilities. When an electric or magnetic field is applied, or a galvanic cell is formed with other metals, the interior flow field will keep moving, which is closely related to the movement of the electric charge and the excitation of the endogenous magnetic field. In this sense, the rich operability of liquid metal machines offer quite a few experimental approaches for the discovery of magnetic monopoles.

The magnetic field described by Maxwell's electromagnetic equations is a passive field, that is, each closed magnetic line of induction does not have a starting point and an end point. If the magnetic monopole exists, the magnetic field will become an active field. Maxwell's equations need to undergo appropriate coordinate transformation and conditional restrictions, which does not violate its correctness. The revised equations are as follows,

$$\nabla \times \boldsymbol{E} = -\frac{\partial \boldsymbol{B}}{\partial t} - \boldsymbol{J}_m \qquad (30)$$

$$\nabla \cdot \boldsymbol{B} = \rho_m \qquad (31)$$

Compared with Maxwell's equations, which originally did not consider the existence of magnetic monopoles, the two differential equations related to the magnetic field have changed. Since the movement of the magnetic monopole may also excite the electric field, the magnetic current density vector $J_m$ is introduced into the Equation (12) to obtain the Equation (30). At the same time, the magnetic monopole can excite the magnetic field by itself, and the divergence of the magnetic field is no longer zero (Equation 31). The equations will show a more symmetrical electromagnetic field excitation form. Under the new category of liquid metal magnetic monopole, many theoretical and experimental works are worthy of pursuing in the coming time.

**4.4 Endogenous Magnet from Rigid, Soft to Fluidic Matters**

From the initial discovery of natural magnetic fields, the manufacture of artificial magnets, to the conclusion of electromagnetic related theories, people have experienced a long exploration in this process. Based on the above systematic interpretation, we outlined **Figure 14** as follows which comparatively illustrates the evolution diagram of various forms of magnets that people have manufactured, i.e. From rigid matter, to soft material until not fluidic magnetism.

**Figure 14a** depicts magnets of various shapes that are widely known and commonly used in daily life. Many important electromagnetic field theories are based on that. **Figure 14b** refers to a soft magnetic substance, which can be attached to a specific substrate. This magnetic material has been applied to various electronic devices, such as sensors and flexible circuits. **Figure 14c** is a composite of magnetic nanoparticles and fluid proposed in recent years. This fluid has no magnetic attraction in static state, and only exhibits magnetism when an external magnetic field is applied. It has a wide range of applications in the fields of medical equipment, magnetic fluid beneficiation, but is still not an ideal all-magnetic fluid. Stepping further, the rotating liquid metal machine or motor as proposed in this work belongs to a new kind of magnetic fluid, or electromagnefluid, as shown in **Figure 14d**. In this category, there is no need to add additional magnetic particles, which is a complete magnetic fluid in the true sense. The discovery of room temperature liquid metal full magnetic fluid will bring new visions and directions for research in diverse fields, which has profound significance for the academic society to further understand magnets and fully use them.

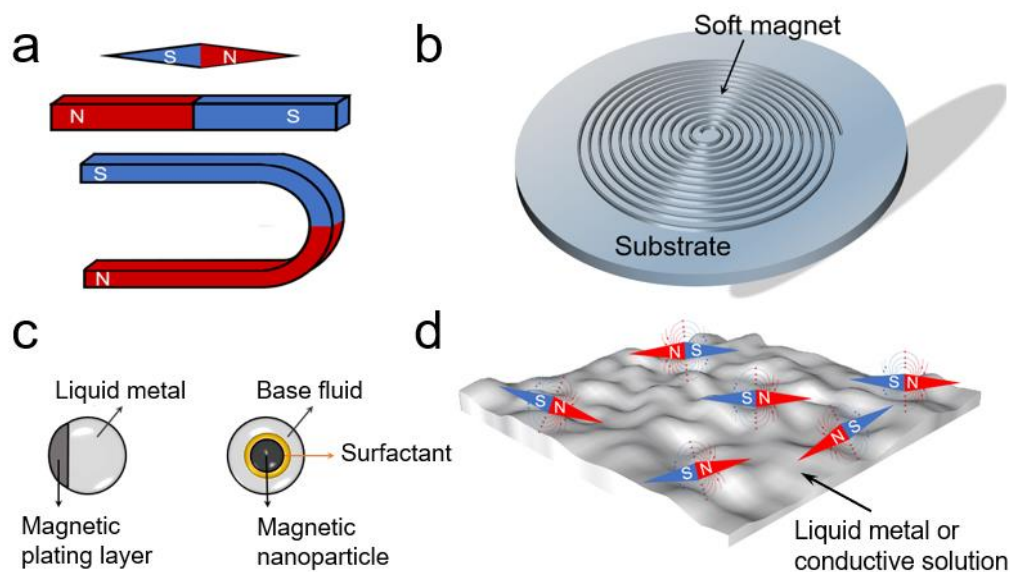

**Figure 14.** Different forms of magnets: (a) Rigid magnet; (b) Soft magnet; (c) Semiliquid magnet; (d) Liquid magnet.

In nature, the north and south poles always exist at the same time. Cut a magnet into two pieces, and each piece will form its own north and south poles, which makes it difficult to achieve the existence of one pole alone. Compared with traditional rigid magnets, the advantage of liquid metal lies in its fluidity, which makes it easier to adjust to different shapes and the restriction on particle movement is reduced. While the surface charges are moving fast, the inside of the fluid is constantly rotating, making the surrounding magnetic field in a state of constant change. If the single charge on the surface of the liquid metal can be manipulated, even controlling the positive and negative features of the charge in a certain position, then the direction of a single magnetic particle can be reversed, it is possible to construct a liquid metal magnetic monopole.

Furthermore, if the direct experimental observation of the magnetic monopole can be achieved, the charge quantification can be better explained. Electrons are particles with no internal structure. They can be reshaped to carry orbital angular momentum (OAM). By coupling the center of mass of the wave function with the interior, the total angular momentum information can be obtained from the vortex beam, and then it can be manipulated artificially[51]. Similarly, if the magnetic moment of the liquid metal micromotor can be measured to obtain momentum information, it may be manipulated and improved microscopically, and it will even be possible to develop technology based on moving magnetic charges to break through the limitations of current charge engineering and create magnetic materials that exhibit richer behavior.

## 5 Conclusion

Liquid metal can interact with sound, light, heat, electricity and magnetism etc. The present study disclosed that as a fluidic conductor, liquid metal machine opens a generalized way to generate transformable endogenous magnet through tuning its interior rotational configurations. This mechanism is rather difficult to implement through rigid materials otherwise. It may change people's basic understanding of the classical magnetism science. A most noteworthy tool enabled from this finding still lies in that such fluidic magnet suggested a new promising way to realize magnetic monopole in reality. And the fundamental routes can be based on either self-fueled liquid metal motor or the electrically controlled spin motion of liquid metal machine inside the electrolyte. Meanwhile, it should also be pointed out that, such liquid metal endogenous magnetism is fundamentally different from the conventional magnetic fluids which was in fact caused by the magnetic particles rather than the fluid itself. This raised important fundamental and practical issues waiting to be addressed in the coming time. It also opens perspective for making new conceptual liquid machines which may find some emerging uses in future basic physics, information technology, smart devices, and advanced functional materials. Overall, the reasonable conjecture as made on the classical mystery whether magnetic monopole exists or not would shed light to innovate future science and engineering.

**Acknowledgments**

This work was partially supported by the National Natural Science Foundation of China under Grants No. 91748206 and No. 51890893.